\documentclass[aps,pre,preprint]{revtex4-1}
\usepackage{graphicx}
\usepackage{amsmath}
\usepackage{subfigure}
\usepackage{url}

\begin{document}

\title{Comparing periodic-orbit theory to perturbation theory in the asymmetric infinite square well}

\author{Todd K. Timberlake}
\email{ttimberlake@berry.edu}
\affiliation{Department of Physics, Astronomy, and Geology, Berry College,
Mount Berry, Georgia 30149-5004}

\date{\today}

\begin{abstract}
An infinite square well with a discontinuous step is one of the simplest systems to exhibit non-Newtonian ray-splitting periodic orbits in the semiclassical limit.  This system is analyzed using both time-independent perturbation theory (PT) and periodic-orbit theory and the approximate formulas for the energy eigenvalues derived from these two approaches are compared.  The periodic orbits of the system can be divided into classes according to how many times they reflect from the potential step.  Different classes of orbits contribute to different orders of PT.  The dominant term in the second-order PT correction is due to non-Newtonian orbits that reflect from the step exactly once.  In the limit in which PT converges the periodic-orbit theory results agree with those of PT, but outside of this limit the periodic-orbit theory gives much more accurate results for energies above the potential step.
\end{abstract}

\maketitle

\section{Introdution}
\label{intro}

Periodic-orbit theory is one of the most interesting developments in the study of the relationship between classical and quantum mechanics \cite{Gutzwiller}.  The centerpiece of periodic-orbit theory is the Gutzwiller trace formula which relates the quantum density of states to properties of the periodic orbits in the classical system.  In general the trace formula gives only approximate results, becoming exact in the limit $\hbar \to 0$.  However, for some systems the periodic-orbit theory is exact even for finite $\hbar$ \cite{Bhullar2006}.  Periodic-orbit theory has found particularly fruitful application in the study of quantum systems with chaotic classical counterpart \cite{Haake2001,Gutzwiller,Reichl2004,Stockmann1999}.

Recently periodic-orbit theory has been extended to the case of ray-splitting systems \cite{Couchman1992}.  In most systems the wavelength of a quantum particle in the semiclassical limit ($\hbar \to 0$) is small compared to all relevant length scales in the classical system.  In this case the quantum wave equations reduce to ray equations and the particle obeys Newtonian mechanics.  In ray-splitting systems, however, the potential changes significantly even on length scales that are small compared to the wavelength of the quantum particle in the semiclassical limit.  This will occur if there is a discontinuous change in the potential within the region accessible to the particle.  In the semiclassical limit of a ray-splitting system a particle will follow Newtonian mechanics everywhere except at the discontinuous boundary.  At the discontinuity the particle may be transmitted across the boundary or reflected from it, so the ray splits into two parts.  The non-Newtonian periodic orbits that result from this ray-splitting can influence the quantum dynamics.  Recent computational and experimental studies of ray-splitting systems have revealed signatures of non-Newtonian periodic orbits in the Fourier transform of the quantum density of states \cite{Blumel1996,Blumel1996b,Sirko1997,Kohler1997,Bauch1998,Schafer2001}, the distribution of level spacings \cite{Oerter1996,Blumel1996b, Timberlake2009}, and the scarring of energy eigenstates \cite{Blumel1996,Blumel1996b,Kohler1997,Schafer2001}.

Periodic-orbit theory is typically applied to systems with chaotic classical dynamics.  The quantum versions of these systems are generally not amenable to exact solution, or even approximation methods like perturbation theory, and must be studied numerically.  Ray-splitting systems are interesting in this regard because their classical dynamics may exhibit some properties of chaos (like an exponential proliferation of periodic orbits with increasing period) but perturbation theory may still be applied to the quantum dynamics of the system (at least for certain parameter regimes).  For some of these systems exact formulas have been found which give the quantum energy eigenstates as a sum over the classical (Newtonian and non-Newtonian) periodic orbits.  These cases allow for a direct analytical comparison between the results of perturbation theory and those of periodic-orbit theory.  The goal of this study is to carry out this comparison for the simplest possible system: the asymmetric infinite square well (AISW).  

The AISW consists of an infinite square well of width $2a$ with a discontinuous step of height $V_0$ at the center of the well \cite{Doncheski2000,Gilbert2005,Jensen2005}.  The potential energy function is
\begin{equation}
V(x) = \left\{
	\begin{array}{ll}
	\infty, & |x| \geq a\\
	0, & -a<x \leq 0\\
	V_{0}, & 0<x<a.
	\end{array}
	\right.
	\end{equation}
A plane wave with energy $E>V_0$ incident on the boundary at $x=0$ may be reflected with probability $r^2$ or transmitted with probability $1-r^2$, where
\begin{equation}
\label{reflection}
 r = \frac{1-\sqrt{1-V_0/E}}{1+\sqrt{1-V_0/E}}. 
 \end{equation}
Note that $r$ does not depend on $\hbar$ so these non-Newtonian reflections persist in the semiclassical limit and the classical dynamics contains non-Newtonian periodic orbits such as those shown in Figure \ref{aisw}.  The orbit $N$ is the Newtonian orbit that moves back and forth between the hard walls at $x=\pm a$.  The orbit $L$ is a non-Newtonian orbit that reflects when it reaches the boundary at $x=0$ from the left, so this orbit is confined to the left side of the well.  The orbit $R$ is confined to the right side of the well, reflecting when it reaches the boundary at $x=0$.  These three basic orbits can be combined to form an infinite variety of other periodic orbits.  Non-Newtonian orbits are also possible for $E < V_0$, but in this case the orbits are ``ghost orbits'' that explore the classically-forbidden right side of the well.  For perturbation theory to be valid the energy of the particle must be greater than $V_0$, so there will be no need to consider the case $E<V_0$ in this paper.

Section \ref{pt} presents the results of a second-order (Rayleigh-Schr\"{o}dinger) time-independent perturbation theory (PT) analysis of the AISW.  Section \ref{po} details the application of periodic-orbit theory to this system, with the goal of providing an approximate formula for the energies that is comparable in accuracy to the second-order PT results.  The application of periodic-orbit theory to the AISW clearly shows that periodic orbits with different numbers of reflections at $x=0$ contribute to different orders of PT.  The second-order PT correction is due predominantly to non-Newtonian orbits that have a single reflection from the potential step.    Section \ref{comparison} presents a comparison of the results of PT and periodic-orbit theory to the exact energy eigenvalues.  The approximation derived from periodic-orbit theory is shown to be more accurate than that derived from PT.  Section \ref{conclusion} provides a summary and discussion of the results.  Most of the details for the PT and periodic-orbit theory calculations are given in Appendices at the end of the paper.

\section{Perturbation theory}
\label{pt}

Standard (Rayleigh-Schr\"{o}dinger) time-independent perturbation theory proceeds by writing the full Hamiltonian for the system as $H= H_0+H'$ where $H_0$ is the Hamiltonian of the ``unperturbed'' system and $H'$ is the perturbation.  For the AISW the unperturbed system is simply an infinite square well with potential function
\begin{equation}
\label{SWpotential}
V(x) = \left\{ \begin{array}{ll}
	\infty, & |x| \geq a \\
	0, & |x| < a
	\end{array} \right.
\end{equation}
where the width of the well is $2a$.  This system is discussed in almost any textbook on elementary quantum mechanics \cite{Griffiths2005}.  The energy eigenvalues of this system are
\begin{equation}
\label{SWenergies}
E_n^{(0)} = \frac{\pi^2\hbar^2n^2}{8ma^2}
\end{equation}
where $m$ is the mass of the particle in the well.  The wave function for the energy eigenstates are
\begin{equation}
\label{SWwavefunctions}
\psi_n^{(0)} (x) = \frac{1}{\sqrt{a}} \sin\left( \frac{n\pi (x+a)}{2a} \right).
\end{equation}

The perturbation for the AISW is then
\begin{equation}
\label{PTpotential}
H' = \left\{ \begin{array}{ll}
	0, & x \leq 0\\
	V_0, & 0 < x < a\\
	0, & x \geq a
	\end{array} \right.
\end{equation}
where $V_0$ is the height of the potential step inside the well.  The infinite square well has no degeneracies, so non-degenerate PT can be applied to find approximations for the energy eigenvalues of the AISW.  The first order correction is
\begin{equation}
\label{FOPT}
E^{(1)}_n =\langle \psi_n^{(0)} | V_p | \psi_n^{(0)} \rangle  =  V_0 \int_0^a \sin^2 \left( \frac{n\pi(x+a)}{2a}\right) dx = \frac{V_0}{2}.
\end{equation}
This correction simply shifts each energy eigenvalue upward by half the height of the potential step.  The second order correction is given by
\begin{equation}
\label{SOPTeq1}
E^{(2)}_n  =  \sum_{k \neq n} \frac{|\langle \psi^{(0)}_k|V_p|\psi^{(0)}_n \rangle|^2}{E^{(0)}_n-E^{(0)}_k}.
\end{equation}
The second order correction is evaluated in App. \ref{pt2} and the result is
\begin{equation}
\label{SOPTfinal}
E^{(2)}_n = \frac{\gamma_n m a^2 V_0^2}{2\pi^2 \hbar^2 n^2} + O\left( \frac{1}{n^3} \right)
\end{equation}
where
\begin{equation}
\label{gamma}
\gamma_n = \left\{
   	\begin{array}{ll}
	3, & \mbox{$n$ is even}\\
	-1, & \mbox{$n$ is odd}\\
	\end{array} \right.
\end{equation}
and $O(g(n))$ indicates a function $f(n)$ that is less than some constant times $g(n)$ for all $n$ greater than some value $n=C$.  So to second order the energies of the AISW are given by
\begin{equation}
\label{PTresult}
E_n = \frac{\pi^2 \hbar^2 n^2}{8ma^2}+\frac{V_{0}}{2} + \frac{\gamma_n m a^2 V_0^2}{2\pi^2 \hbar^2 n^2} + O\left( \frac{1}{n^3} \right).
\end{equation}
The $O(1/n^3)$ terms can be ignored for sufficiently large $n$.

To simplify the notation it is helpful to introduce a dimensionless constant
\begin{equation}
\label{alpha}
\alpha = \frac{ma^2 V_0}{\hbar^2}.
\end{equation}
Note that this constant provides information about the size of the potential step $V_0$ relative to the ground state energy of the infinite square well $E_1^{(0)}$, since
\begin{equation}
\label{ratio}
\frac{V_0}{E_1^{(0)}} = \frac{8ma^2V_0}{\pi^2\hbar^2} = \frac{8\alpha}{\pi^2}.
\end{equation}
In terms of this dimensionless constant the AISW energies are approximately given by 
\begin{equation}
\label{PTfinal}
E_n = E_1^{(0)} \left( n^2 + \frac{4\alpha}{\pi^2} + \frac{4\gamma_n \alpha^2}{\pi^4 n^2}  \right) + O\left(\frac{1}{n^3}\right).
\end{equation}

Before any attempt is made to use the results of PT one must carefully consider whether or not the perturbation expansion will converge.  The requirement for the rapid convergence of the perturbation expansion is \cite{Merzbacher1970}
\begin{equation}
\label{converge}
\left| \frac{\langle \psi_k^{(0)} | V_p | \psi_n^{(0)} \rangle}{E_n^{(0)}-E_k^{(0)}} \right| = \frac{8\alpha}{\pi^3} \left| \frac{(-1)^n (k-n)-k-n}{(k^2-n^2)^2}\right| << 1.
\end{equation}
It is clear that the left side of Eq. \ref{converge} will be largest when $k=n\pm 1$, in which case
\begin{eqnarray}
\label{converge2}
\left| \frac{\langle \psi_{n\pm 1}^{(0)} | V_p | \psi_n^{(0)} \rangle}{E_n^{(0)}-E_{n\pm 1}^{(0)}} \right| & = & \frac{8\alpha}{\pi^3}  \frac{2n\pm 1 \mp (-1)^n}{(2n\pm 1)^2} \nonumber \\
& = & \frac{4\alpha}{\pi^3 n} + O\left(\frac{1}{n^2}\right).
\end{eqnarray}
So the condition for the convergence of PT reduces to
\begin{equation}
\label{PTcriterion}
\frac{\alpha}{n} << 1.
\end{equation}

\section{Periodic Orbit Theory}
\label{po}

Dabaghian and Jensen derived an exact formula for the energy eigenvalues for the AISW in terms of an infinite sum over the periodic orbits of the classical system \cite{Jensen2005}. The procedure they used does not converge for $E<V_0$ \cite{BlumelComment}, but a convergent semiclassical formula for $E<V_0$ can be obtained by accounting for ghost orbits (orbits that exist in the right side of the well, which is forbidden in Newtonian mechanics) \cite{BlumelComment,Bhullar2006}.  The analysis in this paper is limited to large $n$ for which $E_n > V_0$, so the effect of ghost orbits will be ignored and the simpler formula of Ref. \onlinecite{Jensen2005} will be used.

Dabghian and Jensen give their semiclassical formula in terms of the discrete quantum actions $S_n$.  These quantum actions are related to the energy eigenvales of the system via the definition of the classical action length for a particle with energy $E$ moving across the well:
\begin{equation}
\label{ActionLength}
S(E) = a\sqrt{2mE} + a\sqrt{2m(E-V_0)}.
\end{equation}
This equation can be inverted to find the energy as a function of action
\begin{equation}
\label{EofS}
E(S) = \frac{(S^2 + 2ma^2V_0)^2}{8ma^2S^2}
\end{equation}
and the quantum energy eigenvalues are then given by $E(S_n)$.

The semiclassical formula for the reduced quantum action ($s_n = S_n/\hbar$) associated with the $n$th energy eigenstate is \cite{Jensen2005}:
\begin{equation}
\label{POaction}
s_n = 2\pi n - \frac{\pi}{2} - \int_{\pi(n-1/2)}^{\pi(n+1/2)} \bar{N}(s) ds - \frac{1}{\pi} \mbox{Im} \sum_{p,\nu} \int_{\pi(n-1/2)}^{\pi(n+1/2)} \frac{A_p^{\nu}}{\nu} e^{i\nu s_p} ds
\end{equation}
where the variable $s$ is the reduced classical action length $s=S/\hbar$.  $\bar{N}(s)$ is the Weyl average for the density of states with $E>V_0$ in the AISW, given by
\begin{equation}
\label{nbar}
\bar{N}(s) = \frac{s}{\pi} - \frac{1}{2}.
\end{equation}
The first integral in Eq. \ref{POaction} can be evaluated and the formula for the reduced actions can be simplified to
\begin{equation}
\label{sn}
s_n = \pi (n - \omega_n)
\end{equation}
where
\begin{equation}
\label{omega}
\omega_n \equiv \frac{1}{\pi^2} \mbox{Im} \sum_{p,\nu} \int_{\pi(n-1/2)}^{\pi(n+1/2)} \frac{A_p^{\nu}}{\nu} e^{i\nu s_p} ds.
\end{equation}
The index $p$ labels all of the fundamental periodic orbits of the system (orbits which are not simply repetitions of other periodic orbits).  The index $\nu$ accounts for repetitions of these orbits.  The classical action of each periodic orbits is given by
\begin{equation}
\label{bigSp}
S_p = n_L 2a\sqrt{2mE} + n_R 2a\sqrt{2m(E-V_0)}
\end{equation}
where $n_L$ and $n_R$ are the number of times the orbits passes back and forth across the left and right sides of the well, respectively.  The quantity $s_p$ which appears in Eq. \ref{omega} is then the reduced classical action for a periodic orbit:
\begin{equation}
\label{sp}
s_p = \frac{S_p}{\hbar} = (n_L+n_R)s + \frac{2\alpha(n_L-n_R)}{s}.
\end{equation}
The factor $A_p$ is a weighting factor given by
\begin{equation}
\label{ap}
A_p = (-1)^{\chi(p)} r^{\sigma(p)} t^{\tau(p)}
\end{equation}
where $\sigma(p)$ counts the number of times the orbit $p$ reflects from the barrier at $x=0$, $\tau(p)$ counts the number of times the orbit transmits through the barrier, and $\chi(p)$ counts the combined number of reflections from the hard walls and right reflections from the boundary (each of which results in a sign change in the particle's wave function).  The reflection coefficient $r$ is found by substituting Eq. \ref{EofS} into Eq. \ref{reflection} to find
\begin{equation}
\label{rs}
r = \frac{2ma^2V_0}{S^2} = \frac{2\alpha}{s^2}
\end{equation}
with $\alpha$ defined as in Eq. \ref{alpha}.  The transmission cofficient $t$ is given by
\begin{equation}
\label{t}
t^2 = 1-r^2 = 1-\frac{4\alpha^2}{s^4}.
\end{equation}
A more detailed discussion of this formula can be found in Ref. \onlinecite{Jensen2005}.

Substituting Eq. \ref{sn} into Eq. \ref{EofS} provides a formula for the energy eigenvalues of the AISW in terms of $n$ and $\omega_n$:
\begin{eqnarray}
\label{POenergy}
E_n = E(\hbar s_n) & = & \frac{\pi^2\hbar^2 n^2}{8ma^2} + \frac{V_0}{2} - \frac{\pi^2\hbar^2 n \omega_n}{4ma^2} + \frac{\pi^2\hbar^2  \omega_n^2}{8ma^2} + \frac{ma^2V_0^2}{2\hbar^2\pi^2 (n - \omega_n)^2} \nonumber \\
 & = & E_1^{(0)} \left( n^2 + \frac{4\alpha}{\pi^2} - 2n\omega_n + \omega_n^2 + \frac{4\alpha^2}{\pi^4 n^2}\right) + O\left(\frac{1}{n^3}\right).
\end{eqnarray}
A comparison of this result with the PT formula (Eq. \ref{PTfinal}) shows that the zeroth and first order PT terms are also present in Eq. \ref{POenergy}. These terms arise from the non-oscillatory part of Eq. \ref{POaction} which is associated with the Weyl average (Eq. \ref{nbar}).  To compare the periodic-orbit theory result to the second-order PT correction it is necessary to evaluate the oscillatory term $\omega_n$ using Eq. \ref{omega}.  This formula involves a sum over all (Newtonian and non-Newtonian) periodic orbits of the classical system, including all repetitions of periodic orbits.  The Newtonian orbit and its repetitions will be addressed first.

\subsection{Newtonian orbits}
\label{newt}

The Newtonian periodic orbit passes back and forth across the entire well from $x=-a$ to $x=a$ without reflecting from the potential discontinuity at $x=0$.  For this orbit $n_L=n_R=1$, and  Eq. \ref{sp} then gives $s_p = 2s$.  Since this orbit reflects once off of each of the two hard walls, but does not reflect from the potential step, it is clear that $\sigma(p)=0$, $\tau(p)=2$, and $\chi(p)=2$.  If the contribution of the Newtonian orbits and its repetitions to $\omega_n$ is designated by $\omega_{n,0}$ then
\begin{eqnarray}
\label{omega0}
\omega_{n,0} & = & \frac{1}{\pi^2} \mbox{Im} \sum_{\nu=1}^{\infty} \frac{1}{\nu} \int_{\pi(n-1/2)}^{\pi(n+1/2)} \left(1-\frac{4\alpha^2}{s^4}\right)^{\nu} e^{i2\nu s} ds \nonumber \\
  & = & \omega_{n,0} = \frac{8 \log(2) \alpha^2}{\pi^6 n^5} + O\left( \frac{1}{n^6} \right).
\end{eqnarray}
The details of this calculation are given in App. \ref{newtonian}.  Now $\omega_n$ appears twice in Eq. \ref{POenergy}: once in a term involving $n\omega_n$ and again in a term involving $\omega_n^2$.  From Eq. \ref{omega0} it is clear that 
\begin{equation}
\label{nomega0}
n\omega_{n,0} \in O\left( \frac{1}{n^4} \right)
\end{equation}
while
\begin{equation}
\label{omega0squared}
\omega_{n,0}^2 \in O\left( \frac{1}{n^{10}} \right).
\end{equation}
Therefore the Newtonian orbit and its repetitions make a contribution to the energy formula in Eq. \ref{POenergy} that is smaller than other terms that have been ignored in that equation.  At the level of approximation given by Eq. \ref{POenergy} the Newtonian orbit can be ignored.

\subsection{Non-Newtonian orbits}
\label{nonnewt}
Since the Newtonian periodic orbit makes no significant contribution to the energy formula in Eq. \ref{POenergy} it is necessary to examine the contributions of non-Newtonian periodic orbits.  It is convenient to divide the non-Newtonian orbits into classes based on how many times each orbits reflects from the potential step at $x=0$.  If $p_k$ designates a periodic orbit with $k$ reflections at the step then $\sigma(p_k)=k$, so the contribution to $\omega_n$ from orbits with $k$ reflections is  
\begin{equation}
\label{omegak}
\omega_{n,k} =  \frac{1}{\pi^2} \mbox{Im} \sum_{p_k} (-1)^{\chi(p_k)} \int_{\pi(n-1/2)}^{\pi(n+1/2)} \left( \frac{2\alpha}{s^2}\right)^k \left(1-\frac{4\alpha^2}{s^4}\right)^{\tau(p_k)/2} e^{i s_{p_k}} ds
\end{equation}
where the sum is taken over all orbits with $k$ reflections at the step.  Note the absence of the sum over $\nu$, which was used to account for repetitions of orbits in Eq. \ref{omega}.  When calculating the contribution of the orbits $p_k$ to $\omega_n$ it is unnecessary to consider repetitions of these orbits, since the repetition of an orbit with $k$ reflections would be an orbit with $2k$ reflections which would belong to a different class.

Examination of Eq. \ref{omegak} reveals the way in which different classes of orbits contribute to different orders of perturbation theory.  The formula for $\omega_{n,k}$ contains a term proportional to $\alpha^k$ and other terms that involve higher powers of $\alpha$.  The $k$th order correction from perturbation theory is always proportional to $\alpha^k$ so periodic orbits with $k$ reflections can only contribute to perturbation theory corrections of $k$th order and higher.  The Newtonian orbit (and its repetitions) can contribute to all orders, but we have seen above that the zeroth and first order corrections (as well as part of the second order correction) come from the Weyl average while the contribution of the Newtonian orbit to the oscillatory term $\omega_n$ has a negligible effect on the second order term.  Single reflection orbits can contribute, in principal, to first order and higher corrections.  Orbits with three or more reflections cannot contribute to the second order correction from perturbation theory, so they need not be considered in comparing Eq. \ref{POenergy} to Eq. \ref{PTfinal}.

A closer look at Eq. \ref{omegak} reveals that two-reflection orbits make only a negligible contribution to the second order PT correction.  The variable of integration, $s$, in Eq. \ref{omegak} is approximately equal to $\pi n$ throughout the range of integration, so
\begin{equation}
\label{omegakOrd}
\omega_{n,k} \in O \left( \frac{1}{n^{2k}} \right) \to n\omega_{n,k}\in O\left(\frac{1}{n^{2k-1}}\right), \omega_{n,k}^2 \in O\left(\frac{1}{n^{4k}}\right).
\end{equation}
So $\omega_{n,2} \in O(1/n^4)$ and the contribution to the energy formula from two-reflection orbits is $O(1/n^3)$.  Since Eq. \ref{PTfinal} already ignores terms of this order the contribution of two-reflection orbits can also be ignored.

These arguments imply that the dominant term from the second order PT correction \emph{must} come from the contribution of single reflection non-Newtonian periodic orbits (in combination with the $4\alpha^2/(\pi^4 n^2)$ term in Eq. \ref{POenergy} which comes from the non-oscillatory part of Eq. \ref{sn}).  This result is shown explicitly in the next section.

\subsection{Single reflection orbits}
\label{single}

Appendix \ref{onereflection} provides a detailed calculation of $\omega_{n,1}$.  The result is
\begin{equation}
\label{omega1}
\omega_{n,1} = (-1)^{n+1} \frac{2\alpha}{n^2\pi^3} \sin\left(\frac{2\alpha}{n\pi}\right) + O\left( \frac{1}{n^4}\right).
\end{equation}
Since the contribution of the Newtonian orbit and its repetitions is $\omega_{n,0} \in O(1/n^5)$ and the contribution of $k$-reflection orbits is $\omega_{n,k} \in O(1/n^{2k})$ as shown above, then
\begin{equation}
\label{omegaOrd}
\omega_{n} = \omega_{n,1} + O\left( \frac{1}{n^4}\right).
\end{equation}
So in comparing Eq. \ref{POenergy} with Eq. \ref{PTfinal} only the contributions to $\omega_n$ from single reflection non-Newtonian orbits need to be considered.  Inserting Eq. \ref{omega1} in place of $\omega_n$ in Eq. \ref{POenergy} gives
\begin{equation}
\label{POresult}
E_n = E_1^{(0)} \left[ n^2 + \frac{4\alpha}{\pi^2} + (-1)^{n}\frac{4\alpha}{n\pi^3}\sin\left( \frac{2\alpha}{n\pi}\right) + \frac{4\alpha^2}{n^2\pi^4} \right] + O\left( \frac{1}{n^3}\right).
\end{equation}

The perturbation theory approximation of Eq. \ref{PTfinal} is only valid if $\alpha/n << 1$.  In this case the sine function in Eq. \ref{POresult} can be approximated as
\begin{equation}
\label{sine}
\sin\left(\frac{2\alpha}{n\pi}\right) = \frac{2\alpha}{n\pi} + O\left( \frac{1}{n^3}\right).
\end{equation}
Inserting this result into Eq. \ref{POresult} and combining like terms gives
\begin{eqnarray}
\label{equal}
E_n & = & E_1^{(0)} \left[ n^2 + \frac{4\alpha}{\pi^2} + (-1)^n \frac{8\alpha^2}{n^2\pi^4} + \frac{4\alpha^2}{n^2\pi^4} \right] + O\left( \frac{1}{n^3}\right) \nonumber \\
 & = & E_1^{(0)} \left( n^2 + \frac{4\alpha}{\pi^2} + \frac{4\gamma_n \alpha^2}{\pi^4 n^2}  \right) + O\left(\frac{1}{n^3}\right)
\end{eqnarray}
with $\gamma$ defined as in Eq. \ref{gamma}.  So the formula for the approximate energies derived from periodic-orbit theory, including only the contributions to $\omega_n$ from single reflection orbits, matches the formula derived from second order PT (Eq. \ref{PTfinal}) in the regime in which PT is valid.

\section{Comparison of Approximations and Exact Energies}
\label{comparison}

Although the approximations given in Eqs. \ref{PTfinal} and \ref{POresult} are equivalent in the limit $\alpha/n << 1$, it is instructive to examine how well each of these approximations performs outside of this limit by comparing these approximations to the exact energy eigenvalues of the AISW.  For $E>V_0$ the energy eigenstates for the AISW are of the form \cite{Doncheski2000}
\begin{equation}
\label{eigenstates}
 \psi(x) = \left\{ 
	\begin{array}{ll}
	A\sin\left[Q(x+a)\right] & \mbox{for} \ -a < x \leq 0 \\
	B\sin\left[q(x-a)\right] & \mbox{for} \ 0 < x < a,
	\end{array} \right. 
	\end{equation}
where $Q \equiv \sqrt{2mE}/\hbar$ and $q \equiv \sqrt{2m(E-V_0)}/\hbar$.  Requiring $\psi$ and $d\psi/dx$ to be continuous at $x=0$ leads to the energy eigenvalue equation:
\begin{equation}
\label{eveqn}
Q \cos(Qa)\sin(qb) + q\cos(qb)\sin(Qa) = 0. 
\end{equation}
The $n$th energy eigenvalue will lie in the open interval $(\hat{E}_n,\hat{E}_{n+1})$ where 
\begin{equation}
\hat{E}_n = E(S)|_{S=\hbar\pi(n-1/2)}
\end{equation}
with $E(S)$ given in Eq. \ref{EofS} \cite{Jensen2005}.  A simple bisection method can be used to rapidly solve Eq. \ref{eveqn} on this open interval to find the value for $E_n$ \cite{numrec}.  This procedure can be automated to find any number of eigenvalues.

\begin{figure}[h]
\begin{center}
\includegraphics[width=5in]{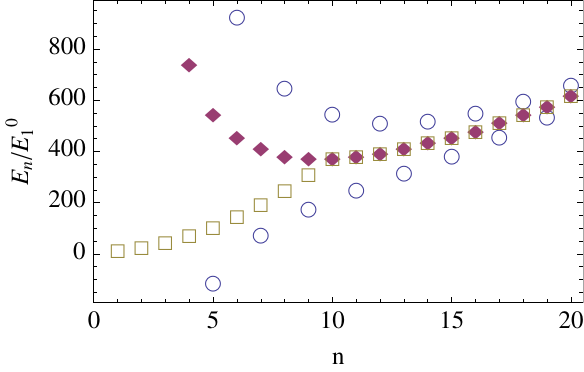}
\end{center}
\caption{\label{compplot} Comparison of approximations with numerically computed eigenvalues.  Open squares show eigenvalues obtained by numerically solving Eq. \ref{eveqn}.  Open circles show the perturbation theory approximation of Eq. \ref{PTfinal}.  Filled diamonds show the periodic-orbit theory approximation of Eq. \ref{POresult}. Parameter values were chosen so that $\alpha=450$.}
\end{figure}

Once the eigenvales are calculated they can be compared with the approximations from periodic-orbit theory and PT.  Figure \ref{compplot} shows the numerically computed energy eigenvalues for the AISW and the approximate values derived from second order PT (Eq. \ref{PTfinal}) and from periodic-orbit theory using only single reflection orbits (Eq. \ref{POresult}).  The parameter values for the data shown are (in scaled units): $a=3$, $V_0=100$, $m=1/2$, and $\hbar=1$ so that the dimensionless quantity $\alpha=450$.  Thus for all of the data shown in Fig. \ref{compplot} $\alpha/n > 20$ and we might not expect either approximation to work well. For the parameter values used in Fig. \ref{compplot} there are 10 eigenstates with $E_n < V_0$.  Note that neither PT nor periodic-orbit theory provide good approximations for $n<10$.  The periodic-orbit theory approximation, though, is quite accurate for $n \geq 10$.  The agreement for $n=10$ is somewhat surprising since the $n=10$ state has an energy below the step in the regime where ghost orbits must be considered and the version of periodic-orbit theory presented above is not strictly valid.  For eigenvalues greater than $V_0$ the periodic-orbit approximation is clearly superior to the PT approximation, even though the two approximations agree in the limit $\alpha/n << 1$.  This general pattern holds true for other sets of parameter values with $\alpha>1$.

\section{Conclusion}
\label{conclusion}

The asymmetric infinite square well, which consists of an infinite square well with a discontinuous potential step at the center, is an unusual system because its energy eigenvalues can be approximated using both perturbation theory (which is usually applied only to systems with near-integrable classical dynamics) and periodic-orbit theory (which is usually applied only to classically chaotic systems).  A comparison of these two approximations reveals that different classes of periodic orbits contribute to different orders of perturbation theory (PT).  The zeroth and first order terms in the PT expansion, as well as part of the second order term, can be derived using only the non-oscillatory term (the Weyl average) from periodic-orbit theory.  A close examination of the oscillatory part of the periodic-orbit theory formula reveals that orbits with $k$ reflections from the potential step can contribute only to $k$th order and higher terms in the PT expansion.  Aside from the terms already accounted for by the Weyl average, the Newtonian orbit and its repetitions (which have no reflections) make no contribution to the zeroth or first order terms and their contribution to the second order term is negligible.  Periodic orbits with a single reflection from the potential step contribute the dominant term in the second order PT correction.

Much of the behavior described above is likely to carry over to a wide variety of other ray-splitting systems.  At high energies the coefficient $r$ for reflection from the potential step will be small.  Because every reflection contributes a factor of $r$ to the weighting factor $A_p$ in the periodic orbit sum (Eq. \ref{POaction}), orbits with fewer reflections will generally make larger contributions to the sum.  Orbits with more reflections will contribute only to higher order terms in the PT expansion.  It also seems that for any ray-splitting system the Newtonian orbit will contribute only to second-order and higher terms in the PT expansion.  This indicates that single-reflection orbits play a particularly important role in providing a semiclassical explanation of the eigenvalue spectrum in ray-splitting systems.

It is also interesting to note that for the AISW periodic-orbit theory (using oscillatory contributions only from single reflection orbits) provides a much more accurate approximation to the energy eigenvalues with $E>V_0$ than does standard perturbation theory.  This raises the possibility of using periodic-orbit theory to find accurate approximations for energy eigenvalues in other ray-splitting systems.  It would also be interesting to compare the periodic-orbit theory approximation with results from other forms of time-independent perturbation theory like the Dalgarno-Lewis method \cite{DalgarnoLewis} or logarithmic perturbation theory \cite{LogPT}.  High-accuracy analytical approximations for the energy eigenvalues of ray-splitting systems could be useful for a variety of applications, such as studying wave packet revivals in these systems.

\appendix

\section{Second Order Perturbation Theory}
\label{pt2}

The second order correction from PT is given by Eq. \ref{SOPTeq1}
where the matrix element in the denominator is
\begin{eqnarray}
\label{PTmatrix}
\langle \psi^{(0)}_k | V_p | \psi^{(0)}_n \rangle & =  & \frac{V_0}{a} \int_0^a \sin\left(\frac{k\pi (x+a)}{2a}\right) \sin\left(\frac{n\pi (x+a)}{2a}\right) \nonumber  \\
 & = & \frac{V_0}{\pi} \left( -\frac{\sin[(k-n)\pi/2]}{k-n} + \frac{\sin[(k+n)\pi/2]}{k+n} \right).
\end{eqnarray}
The absolute value of this matrix element simplifies to
\begin{equation}
\label{PTmatrix2}
|\langle \psi^{(0)}_k | V_p | \psi^{(0)}_n \rangle|^2 = \left\{ 
  	\begin{array}{ll}
	0, & \mbox{$k$ and $n$ are both even or both odd} \\
	\frac{4V_0^2 k^2}{\pi^2 (k^2-n^2)^2}, & \mbox{odd $n$ and even $k$} \\
	\frac{4V_0^2 n^2}{\pi^2 (k^2-n^2)^2}, & \mbox{even $n$ and odd $k$}. \\
	\end{array} \right.
\end{equation}
So the second order PT correction (Eq. \ref{SOPTeq1}) is
\begin{equation}
\label{SOPTeq2}
E^{(2)}_n  =  \left\{ 
 	\begin{array}{ll}
	\frac{8mV_0^2(a+b)^2n^2}{\pi^4 \hbar^2} \sum_{\mbox{odd $k$}} \frac{1}{(n^2-k^2)^3}, & \mbox{even $n$}  \\
	\frac{8mV_0^2(a+b)^2}{\pi^4 \hbar^2} \sum_{\mbox{even $k$}} \frac{k^2}{(n^2-k^2)^3}, & \mbox{odd $n$}.\\
	\end{array} \right. 
\end{equation}
The infinite sums in Eq. \ref{SOPTeq2} can be approximated by first noting that only values of $k$ close to $n$ (but differing from $n$ by an odd number) will make significant contributions to these sums.  If $k=n+2i-1$ then the first sum in Eq. \ref{SOPTeq2} can be written
\begin{eqnarray}
\label{OddSum}
\sum_{\mbox{odd $k$}} \frac{1}{(n^2-k^2)^3} & = & \sum_{i=(1-n/2)}^{\infty} \frac{1}{[(2i-1)(1-2i-2n)]^3} \nonumber \\
 & = & -\frac{1}{8n^3} \left[ \sum_{i=-\infty}^{\infty} \frac{1}{(2i-1)^3} - \frac{3}{2n} \sum_{i=-\infty}^{\infty} \frac{1}{(2i-1)^2} \right] + O\left(\frac{1}{n^5}\right).
\end{eqnarray}
If $n$ is large then the $O(1/n^5)$ terms can be ignored.  Now
\begin{equation}
\label{SumCubes}
\sum_{i=-\infty}^{\infty} \frac{1}{(2i-1)^3} = 0
\end{equation}
since each positive term in the sum is cancelled by a corresponding negative term of equal absolute value.  The other sum in Eq. \ref{OddSum} is
\begin{equation}
\label{SumSquares}
\sum_{i=-\infty}^{\infty} \frac{1}{(2i-1)^2} = 2\sum_{i=1}^{\infty} \frac{1}{(2i-1)^2} = \frac{\pi^2}{4}.
\end{equation}
Inserting the results from Eq. \ref{SumCubes} and \ref{SumSquares} into Eq. \ref{OddSum} gives
\begin{equation}
\label{OddSum2}
\sum_{\mbox{odd $k$}} \frac{1}{(n^2-k^2)^3}  =  \frac{3\pi^2}{64n^4} + O\left(\frac{1}{n^5}\right).
\end{equation}
Inserting this result into Eq. \ref{SOPTeq2} shows that for even $n$
\begin{equation}
\label{SOPTodd}
E^{(2)}_n = \frac{3m a^2 V_0^2}{2\pi^2 \hbar^2 n^2} + O\left(\frac{1}{n^3}\right).
\end{equation}

The sum that appears in Eq. \ref{SOPTeq2} for odd $n$ can be approximated (using $k=n+2i-1$) as
\begin{eqnarray}
\label{EvenSum}
\sum_{\mbox{even $k$}} \frac{k^2}{(n^2-k^2)^3} & = & \sum_{i=(1-n/2)}^{\infty} \frac{n^2+2n(2i-1)+(2i-1)^2}{[(2i-1)(1-2i-2n)]^3} \nonumber \\
& = & n^2 \sum_{i=(1-n/2)}^{\infty} \frac{1}{[(2i-1)(1-2i-2n)]^3} \nonumber \\
 & &  + 2n \sum_{i=-\infty}^{\infty} \frac{1}{(2i-1)^2(1-2i-2n)^3} + O\left(\frac{1}{n^3}\right).
\end{eqnarray}
The first sum on the right hand side of Eq. \ref{EvenSum} is identical to the sum in Eq. \ref{OddSum} and thus it evaluates to the result given in Eq. \ref{OddSum2}.  The second sum on the right hand side of Eq. \ref{EvenSum} can be expanded in powers of $1/n$ and if only the lowest order term is kept the result is
\begin{eqnarray}
\label{EvenSum2}
\sum_{\mbox{even $k$}} \frac{k^2}{(n^2-k^2)^3} & = &  \frac{3\pi^2}{64n^2} - \frac{1}{4n^2} \sum_{i=-\infty}^{\infty} \frac{1}{(2i-1)^2} + O\left(\frac{1}{n^3}\right) \nonumber \\
 & = & -\frac{\pi^2}{64n^2} + O\left(\frac{1}{n^3}\right).
 \end{eqnarray}
Inserting this result into Eq. \ref{SOPTeq2} shows that for odd $n$
\begin{equation}
\label{SOPTeven}
E^2_n = -\frac{m a^2 V_0^2}{2\pi^2 \hbar^2 n^2} + O\left(\frac{1}{n^3}\right).
\end{equation}
Combining the results for even and odd $n$ provides the result given in Eq. \ref{SOPTfinal} above.

\section{Periodic-Orbit Theory Approximations}

\subsection{Newtonian Orbits}
\label{newtonian}
The contribution of the Newtonian orbit and its repetitions to $\omega_n$ is given by 
\begin{eqnarray}
\label{omega0a}
\omega_{n,0} & = & \frac{1}{\pi^2} \mbox{Im} \sum_{\nu=1}^{\infty} \frac{1}{\nu} \int_{\pi(n-1/2)}^{\pi(n+1/2)} \left(1-\frac{4\alpha^2}{s^4}\right)^{\nu} e^{i2\nu s} ds \nonumber \\
  & = & \frac{1}{\pi^2} \mbox{Im} \sum_{\nu=1}^{\infty} \frac{1}{\nu} \left( \int_{\pi(n-1/2)}^{\pi(n+1/2)} e^{i2\nu s} ds - 4\nu \alpha^2 \int_{\pi(n-1/2)}^{\pi(n+1/2)} \frac{e^{i2\nu s}}{s^4} ds \right) + O\left( \frac{1}{n^8}\right).
\end{eqnarray}
The first integral in the right hand side of Eq. \ref{omega0a} is easily evaluated: 
\begin{equation}
\label{int1}
\int_{\pi(n-1/2)}^{\pi(n+1/2)} e^{i2\nu s} ds = \frac{e^{i2\nu \pi n} \sin(\nu \pi)}{\nu} = 0.
\end{equation}
The second integral in the right hand side of Eq. \ref{omega0a} can be approximated by changing the variable of integration $x = s -n\pi$ and expanding the integrand in powers of $1/n$ to find
\begin{eqnarray}
\label{int2}
\int_{\pi(n-1/2)}^{\pi(n+1/2)} \frac{e^{i2\nu s}}{s^4} ds & = & \int_{-\pi/2}^{\pi/2} \frac{e^{i2\nu(n\pi + x)}}{(n\pi+x)^4} dx \nonumber \\
 & = & \frac{1}{\pi^4 n^4} \int_{-\pi/2}^{\pi/2} e^{i2\nu x} \left( 1 - \frac{4x}{n\pi} \right) dx + O\left( \frac{1}{n^6} \right).
\end{eqnarray}
The integral on the right hand side of Eq. \ref{int2} can then be split into two parts.  The first part evaluates to zero, since it is equivalent to the integral in Eq. \ref{int1} with $n=0$.  The second part involves the integral
\begin{equation}
\label{int3}
\int_{-\pi/2}^{\pi/2} xe^{i2\nu x} dx = \frac{i}{2\nu^2} \left( \sin(\nu\pi)-\pi\nu \cos(\nu\pi)\right) = \frac{i\pi(-1)^{\nu+1}}{2\nu}.
\end{equation}
Substituting these results back into Eq. \ref{omega0a} gives
\begin{equation}
\label{omega0b}
\omega_{n,0} = \frac{8\alpha^2}{\pi^6 n^5} \sum_{\nu=1}^{\infty} \frac{(-1)^{\nu+1}}{\nu} + O\left( \frac{1}{n^6} \right).
\end{equation}
The infinite sum in Eq. \ref{omega0b} can be evaluated:
\begin{equation}
\label{sum0}
\sum_{\nu=1}^{\infty} \frac{(-1)^{\nu+1}}{\nu} = \log(2)
\end{equation}
and substituting this result into Eq. \ref{omega0b} gives the result shown in Eq. \ref{omega0} above.

\subsection{Single Reflection Orbits}
\label{onereflection}
\begin{figure}[h]
\begin{center}
\includegraphics[width=4in]{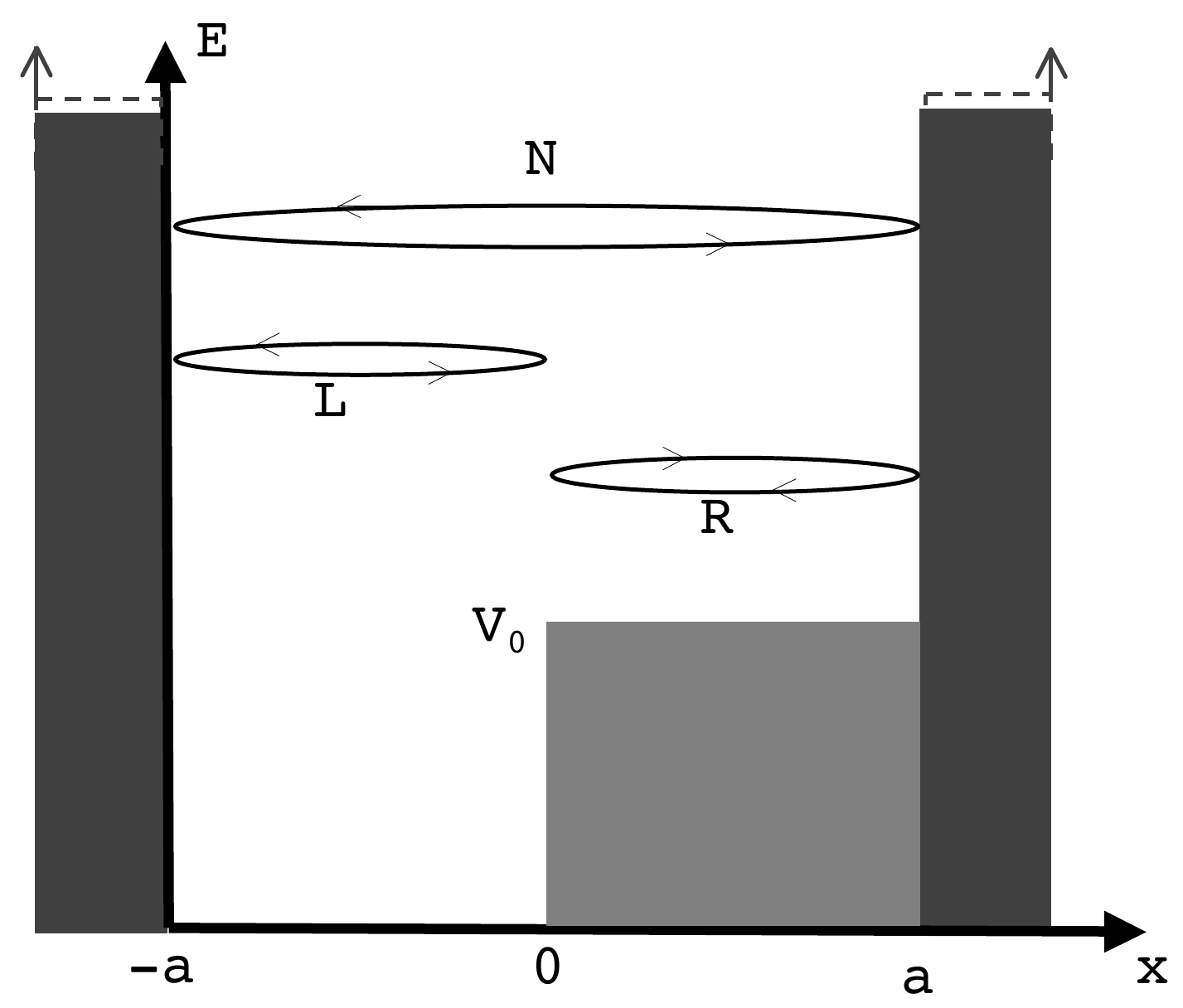}
\end{center}
\caption{\label{aisw} Periodic orbits in the asymmetric infinite square well.  The Newtonian orbit $N$ bounces back and forth between the infinite walls at $x=\pm a$.  The non-Newtonian orbits $L$ and $R$ reflect when they reach the discontinuity in the potential at $x=0$ and are confined to the left and right sides of the well, respectively.}
\end{figure}

Periodic orbits in the AISW can be represented as sequences of the letters $L$ and $R$, where $L$ represents a back-and-forth motion across the left side of the well and $R$ indicates back-and-forth motion across the right side of the well (Fig. \ref{aisw}).  The Newtonian orbit is represented by $LR$ (or $RL$) since it involves motion on one side of the well immediately followed by motion on the other side.  Examples of orbits with a single reflection are: $L$, $R$, $LRL$, $RRL$, $LRLLRLR$, etc.  The cyclic permutation of the symbols in the representation of a periodic orbit just produces a different representation of the \emph{same} orbit (but starting at a different point in the cycle).  Therefore all cyclic permutations of symbols are considered equivalent.  For example, $LRL$ and $RLL$ are really the same orbit.  Each group of equivalent symbol sequences defines a \emph{necklace} (a name chosen to conjure the image of arranging the symbols in a circle).  A particular sequence can be chosen to represent each necklace.

The symbolic necklaces for periodic orbits with exactly one reflection fall into two groups.  The first group consists of necklaces that can be represented by some number of repetitions of the sequence $LR$ followed by a single $L$.  Table \ref{Ltable} lists several of the necklaces in this group along with the values of $n_L+n_R$, $\tau(p)$, and $\chi(p)$ for each necklace.  The bottom row of the table provides general formulas for the $j$th necklace in the group.    The contribution of this group to $\omega_{n,1}$ is designated $\omega_{n,1L}$ and is given by
\begin{eqnarray}
\label{omega1La}
\omega_{n,1L} & = & \frac{2\alpha}{\pi^2} \mbox{Im} \sum_{j=1}^{\infty} (-1)^{2j-1} \int_{\pi(n-1/2)}^{\pi(n+1/2)} \frac{1}{s^2} \left( 1-\frac{4\alpha^2}{s^4} \right)^{j-1} e^{i[(2j-1)s+2\alpha/s]} ds \nonumber \\
 & = & -\frac{2\alpha}{\pi^2} \sum_{j=1}^{\infty} \mbox{Im} \int_{\pi(n-1/2)}^{\pi(n+1/2)} \frac{e^{i(2j-1)s} e^{i2\alpha/s}}{s^2}   ds + O\left(\frac{1}{n^6}\right).
\end{eqnarray}
This integral can be written in terms of the variable $x=s-n\pi$ to give
\begin{equation}
\label{omega1Lb}
\omega_{n,1L}  =  -\frac{2\alpha}{\pi^2}  \sum_{j=1}^{\infty} \mbox{Im} \ e^{i(2j-1)n\pi} \int_{-\pi/2}^{\pi/2} \frac{1}{(n\pi+x)^2} e^{i(2j-1)x} e^{\frac{i2\alpha}{n\pi+x}} dx + O\left(\frac{1}{n^6}\right).
\end{equation}
Note that 
\begin{equation}
\label{euler1}
e^{i(2j-1)n\pi} = (-1)^n
\end{equation}
and two of the factors in the integrand of Eq. \ref{omega1Lb} can be expanded in powers of $1/n$ to give
\begin{equation}
\label{expapprox}
e^{\frac{i2\alpha}{n\pi+x}} = e^{i2\alpha/(n\pi)} \left( 1 - \frac{i2\alpha x}{n^2\pi^2}\right) + O\left(\frac{1}{n^3}\right)
\end{equation}
and
\begin{equation}
\label{ratapprox}
\frac{1}{(n\pi+x)^2} = \frac{1}{n^2\pi^2} \left( 1 - \frac{2x}{n\pi}\right) + O\left(\frac{1}{n^4}\right).
\end{equation}
Substituting these results into Eq. \ref{omega1Lb} gives
\begin{equation}
\label{omega1Lc}
\omega_{n,1L} = (-1)^{n+1} \frac{2\alpha}{n^2\pi^4} \mbox{Im} \ e^{i2\alpha/(n\pi)} \sum_{j=1}^{\infty} \int_{-\pi/2}^{\pi/2} \left( 1 - \frac{(2n\pi + i2\alpha)x}{n^2\pi^2} \right) e^{i(2j-1)x} dx + O\left(\frac{1}{n^5}\right).
\end{equation}
The integral in Eq. \ref{omega1Lc} can be evaluated to give
\begin{eqnarray}
\label{intL}
\int_{-\pi/2}^{\pi/2} \left( 1 - \frac{(2n\pi + i2\alpha)x}{n^2\pi^2} \right) e^{i(2j-1)x} dx & = & \int_{-\pi/2}^{\pi/2} e^{i(2j-1)x} dx - \frac{2n\pi + i2\alpha}{n^2\pi^2} \int_{-\pi/2}^{\pi/2} x e^{i(2j-1)x} dx \nonumber \\
 & = & -\frac{2\cos(j\pi)}{2j-1} - \frac{i2n\pi-2\alpha}{n^2\pi^2} \frac{2\cos(j\pi) + (2j-1)\pi\sin(j\pi)}{(2j-1)^2} \nonumber \\
  & = & (-1)^{j+1} \left(\frac{2}{2j-1} + \frac{4i}{n\pi(2j-1)^2}\right) + O\left(\frac{1}{n^2}\right)
\end{eqnarray}
since $\cos(j\pi)=(-1)^j$ and $\sin(j\pi) = 0$.  The exponential factor in front of the sum in Eq. \ref{omega1Lc} can be expanded using Euler's formula:
\begin{equation}
\label{euler2}
e^{i2\alpha/(n\pi)} = \cos\left(\frac{2\alpha}{n\pi}\right) + i\sin\left(\frac{2\alpha}{n\pi}\right).
\end{equation}
Substituting Eqs. \ref{intL} and \ref{euler2} into Eq. \ref{omega1Lc} gives
\begin{equation}
\label{omega1Ld}
\omega_{n,1L} = \frac{2\alpha (-1)^{n+1}}{n^2\pi^4} \left[ 2\sin\left(\frac{2\alpha}{n\pi}\right) \sum_{j=1}^{\infty} \frac{(-1)^{j+1}}{2j-1} + \frac{4}{n\pi} \cos\left( \frac{2\alpha}{n\pi}\right) \sum_{j=1}^{\infty} \frac{(-1)^{j+1}}{(2j-1)^2} \right] + O\left(\frac{1}{n^4}\right).
\end{equation}
Note that
\begin{equation}
\label{sumL}
\sum_{j=1}^{\infty} \frac{(-1)^{j+1}}{2j-1} = \frac{\pi}{4}
\end{equation}
and
\begin{equation}
\label{catalan}
\sum_{j=1}^{\infty} \frac{(-1)^{j+1}}{(2j-1)^2} = G
\end{equation}
where $G \approx 0.916$ is Catalan's constant.  Inserting the results for these sums into Eq. \ref{omega1Ld} gives
\begin{equation}
\label{omega1Lf}
\omega_{n,1L} = \frac{2\alpha (-1)^{n+1}}{n^2\pi^4} \left[ \frac{\pi}{2}\sin\left(\frac{2\alpha}{n\pi}\right) + \frac{4G}{n\pi} \cos\left(\frac{2\alpha}{n\pi}\right) \right] + O\left( \frac{1}{n^4}\right).
\end{equation}

The necklaces in the second group of single reflection orbits can be represented by repetitions of the sequence $RL$ followed by a single $R$.  Table \ref{Rtable} provides values of $n_L+n_R$, $\tau(p)$, and $\chi(p)$ for several of these necklaces as well as general formulas for the $j$th necklace in this group.  The contribution to $\omega_{n,1}$ from this group of orbits is designated $\omega_{n,1R}$ and is given by
\begin{equation}
\label{omega1Ra}
\omega_{n,1R} = \frac{2\alpha}{\pi^2} \mbox{Im} \sum_{j=1}^{\infty} (-1)^{2j} \int_{\pi(n-1/2)}^{\pi(n+1/2)} \frac{1}{s^2} \left(1 - \frac{4\alpha^2}{s^4} \right)^{j-1} e^{i[(2j-1)s-2\alpha/s]} ds.
\end{equation}
Note that Eq. \ref{omega1Ra} is identical to Eq. \ref{omega1La} except for the change $\alpha \to -\alpha$.  So $\omega_{n,1R}$ can be evaluated by simply changing $\alpha \to -\alpha$ in Eq. \ref{omega1Lf}.  The result is
\begin{equation}
\label{omega1Rf}
\omega_{n,1R} = \frac{2\alpha (-1)^{n+1}}{n^2\pi^4} \left[ \frac{\pi}{2}\sin\left(\frac{2\alpha}{n\pi}\right) - \frac{4G}{n\pi} \cos\left(\frac{2\alpha}{n\pi}\right) \right] + O\left( \frac{1}{n^4}\right).
\end{equation}

The contribution to $\omega_n$ from all of the single reflection orbits is simply the sum of the contributions from the two groups: $\omega_{n,1} = \omega_{n,1L} + \omega_{n,1R}$.  Adding Eqs. \ref{omega1Lf} and \ref{omega1Rf} produces the result given in Eq. \ref{omega1}.

\newpage

\newpage

\begin{table}[h]
\begin{center}
\begin{tabular}{|c|c|c|c|c|}\hline
\textbf{Necklace} & $n_L + n_R$ & $\tau(p)$ & $\chi(p)$ \\ \hline
L & 1 & 0 & 1\\ \hline
LRL=(LR)+L  & 3 &  2 & 3 \\ \hline
LRLRL=2(LR)+L& 5 &  4 & 5 \\ \hline
\vdots & \vdots & \vdots & \vdots \\ \hline
$(j-1)\times$(LR)+L & $2j-1$ & $2j-2$ & $2j-1$ \\ \hline
\end{tabular}
\end{center}
\caption{\label{Ltable} Parameters for periodic orbits with $\sigma(p)=1$ and $n_L-n_R=1$.}
\end{table}

\begin{table}[h]
\begin{center}
\begin{tabular}{|c|c|c|c|c|}\hline
\textbf{Necklace} & $n_L + n_R$ & $\tau(p)$ & $\chi(p)$ \\ \hline
R & 1 & 0 & 2\\ \hline
RLR = (RL)+R  & 3 &  2 & 4 \\ \hline
RLRLR = 2(RL)+R& 5 &  4 & 6 \\ \hline
\vdots & \vdots & \vdots & \vdots \\ \hline
$(j-1)\times$(RL)+R & $2j-1$ & $2j-2$ & $2j$ \\ \hline
\end{tabular}
\end{center}
\caption{\label{Rtable} Parameters for periodic orbits with $\sigma(p)=1$ and $n_L-n_R=-1$.}
\end{table}


\begin{thebibliography}{10}%
\makeatletter
\providecommand \@ifxundefined [1]{%
 \ifx #1\undefined \expandafter \@firstoftwo
 \else \expandafter \@secondoftwo
\fi
}%
\providecommand \@ifnum [1]{%
 \ifnum #1\expandafter \@firstoftwo
 \else \expandafter \@secondoftwo
\fi
}%
\providecommand \enquote [1]{``#1''}%
\providecommand \bibnamefont  [1]{#1}%
\providecommand \bibfnamefont [1]{#1}%
\providecommand \citenamefont [1]{#1}%
\providecommand\href[0]{\@sanitize\@href}%
\providecommand\@href[1]{\endgroup\@@startlink{#1}\endgroup\@@href}%
\providecommand\@@href[1]{#1\@@endlink}%
\providecommand \@sanitize [0]{\begingroup\catcode`\&12\catcode`\#12\relax}%
\@ifxundefined \pdfoutput {\@firstoftwo}{%
 \@ifnum{\z@=\pdfoutput}{\@firstoftwo}{\@secondoftwo}%
}{%
 \providecommand\@@startlink[1]{\leavevmode}%
 \providecommand\@@endlink[0]{}%
}{%
 \providecommand\@@startlink[1]{%
  \leavevmode
  \pdfstartlink
   attr{/Border[0 0 1 ]/H/I/C[0 1 1]}%
   user{/Subtype/Link/A<</Type/Action/S/URI/URI(#1)>>}%
  \relax
 }%
 \providecommand\@@endlink[0]{\pdfendlink}%
}%
\providecommand \url  [0]{\begingroup\@sanitize \@url }%
\providecommand \@url [1]{\endgroup\@href {#1}{\urlprefix}}%
\providecommand \urlprefix [0]{URL }%
\providecommand \Eprint[0]{\href }%
\@ifxundefined \urlstyle {%
  \providecommand \doi [1]{doi:\discretionary{}{}{}#1}%
}{%
  \providecommand \doi [0]{doi:\discretionary{}{}{}\begingroup
  \urlstyle{rm}\Url }%
}%
\providecommand \doibase [0]{http://dx.doi.org/}%
\providecommand \Doi[1]{\href{\doibase#1}}%
\providecommand \bibAnnote [3]{%
  \BibitemShut{#1}%
  \begin{quotation}\noindent
    \textsc{Key:}\ #2\\\textsc{Annotation:}\ #3%
  \end{quotation}%
}%
\providecommand \bibAnnoteFile [2]{%
  \IfFileExists{#2}{\bibAnnote {#1} {#2} {\input{#2}}}{}%
}%
\providecommand \typeout [0]{\immediate \write \m@ne }%
\providecommand \selectlanguage [0]{\@gobble}%
\providecommand \bibinfo [0]{\@secondoftwo}%
\providecommand \bibfield [0]{\@secondoftwo}%
\providecommand \translation [1]{[#1]}%
\providecommand \BibitemOpen[0]{}%
\providecommand \bibitemStop [0]{}%
\providecommand \bibitemNoStop [0]{.\EOS\space}%
\providecommand \EOS [0]{\spacefactor3000\relax}%
\providecommand \BibitemShut [1]{\csname bibitem#1\endcsname}%
\bibitem{Gutzwiller}%
  \BibitemOpen
  \bibfield{author}{%
  \bibinfo {author} {\bibfnamefont{M.~C.}\ \bibnamefont{Gutzwiller}},\ }%
  \emph{\bibinfo {title} {Chaos in classical and quantum mechanics}}\ (\bibinfo
  {publisher} {Springer},\ \bibinfo {address} {New York},\ \bibinfo {year}
  {1990})%
  \bibAnnoteFile{NoStop}{Gutzwiller}%
\bibitem{Bhullar2006}%
  \BibitemOpen
  \bibfield{author}{%
  \bibinfo {author} {\bibfnamefont{A.~S.}\ \bibnamefont{Bhullar}}, \bibinfo
  {author} {\bibfnamefont{R.}~\bibnamefont{Bl\"{u}mel}},\ and\ \bibinfo
  {author} {\bibfnamefont{P.~M.}\ \bibnamefont{Koch}},\ }%
  \bibfield{journal}{%
  \bibinfo {journal} {Phys. Rev. E}\ }%
  \textbf{\bibinfo {volume} {73}},\ \bibinfo {pages} {016211} (\bibinfo {year}
  {2006})%
  \bibAnnoteFile{NoStop}{Bhullar2006}%
\bibitem{Haake2001}%
  \BibitemOpen
  \bibfield{author}{%
  \bibinfo {author} {\bibfnamefont{F.}~\bibnamefont{Haake}},\ }%
  \emph{\bibinfo {title} {Quantum Signatures of Chaos}},\ \bibinfo {edition}
  {2nd}\ ed.\ (\bibinfo {publisher} {Springer},\ \bibinfo {address} {New
  York},\ \bibinfo {year} {2001})%
  \bibAnnoteFile{NoStop}{Haake2001}%
\bibitem{Reichl2004}%
  \BibitemOpen
  \bibfield{author}{%
  \bibinfo {author} {\bibfnamefont{L.~E.}\ \bibnamefont{Reichl}},\ }%
  \emph{\bibinfo {title} {The Transition to Chaos: Conservative Classical
  Systems and Quantum Manifestations}},\ \bibinfo {edition} {2nd}\ ed.\
  (\bibinfo {publisher} {Springer},\ \bibinfo {address} {New York},\ \bibinfo
  {year} {2004})%
  \bibAnnoteFile{NoStop}{Reichl2004}%
\bibitem{Stockmann1999}%
  \BibitemOpen
  \bibfield{author}{%
  \bibinfo {author} {\bibfnamefont{H.-J.}\ \bibnamefont{St\"{o}ckmann}},\ }%
  \emph{\bibinfo {title} {Quantum Chaos: An Introduction}}\ (\bibinfo
  {publisher} {Cambridge University Press},\ \bibinfo {address} {New York},\
  \bibinfo {year} {1999})%
  \bibAnnoteFile{NoStop}{Stockmann1999}%
\bibitem{Couchman1992}%
  \BibitemOpen
  \bibfield{author}{%
  \bibinfo {author} {\bibfnamefont{L.}~\bibnamefont{Couchman}}, \bibinfo
  {author} {\bibfnamefont{E.}~\bibnamefont{Ott}},\ and\ \bibinfo {author}
  {\bibfnamefont{T.~M.}\ \bibnamefont{Antonsen}},\ }%
  \bibfield{journal}{%
  \bibinfo {journal} {Phys. Rev. A}\ }%
  \textbf{\bibinfo {volume} {46}},\ \bibinfo {pages} {6193} (\bibinfo {year}
  {1992})%
  \bibAnnoteFile{NoStop}{Couchman1992}%
\bibitem{Blumel1996}%
  \BibitemOpen
  \bibfield{author}{%
  \bibinfo {author} {\bibfnamefont{R.}~\bibnamefont{Bl\"{u}mel}}, \bibinfo
  {author} {\bibfnamefont{T.~M.}\ \bibnamefont{Antonsen}}, \bibinfo {author}
  {\bibfnamefont{B.}~\bibnamefont{Georgeot}}, \bibinfo {author}
  {\bibfnamefont{E.}~\bibnamefont{Ott}},\ and\ \bibinfo {author}
  {\bibfnamefont{R.~E.}\ \bibnamefont{Prange}},\ }%
  \bibfield{journal}{%
  \bibinfo {journal} {Phys. Rev. Lett.}\ }%
  \textbf{\bibinfo {volume} {76}},\ \bibinfo {pages} {2476} (\bibinfo {year}
  {1996})%
  \bibAnnoteFile{NoStop}{Blumel1996}%
\bibitem{Blumel1996b}%
  \BibitemOpen
  \bibfield{author}{%
  \bibinfo {author} {\bibfnamefont{R.}~\bibnamefont{Bl\"{u}mel}}, \bibinfo
  {author} {\bibfnamefont{T.~M.}\ \bibnamefont{Antonsen}}, \bibinfo {author}
  {\bibfnamefont{B.}~\bibnamefont{Georgeot}}, \bibinfo {author}
  {\bibfnamefont{E.}~\bibnamefont{Ott}},\ and\ \bibinfo {author}
  {\bibfnamefont{R.~E.}\ \bibnamefont{Prange}},\ }%
  \bibfield{journal}{%
  \bibinfo {journal} {Phys. Rev. E}\ }%
  \textbf{\bibinfo {volume} {53}},\ \bibinfo {pages} {3284} (\bibinfo {year}
  {1996})%
  \bibAnnoteFile{NoStop}{Blumel1996b}%
\bibitem{Sirko1997}%
  \BibitemOpen
  \bibfield{author}{%
  \bibinfo {author} {\bibfnamefont{L.}~\bibnamefont{Sirko}}, \bibinfo {author}
  {\bibfnamefont{P.~M.}\ \bibnamefont{Koch}},\ and\ \bibinfo {author}
  {\bibfnamefont{R.}~\bibnamefont{Bl\"{u}mel}},\ }%
  \bibfield{journal}{%
  \bibinfo {journal} {Phys. Rev. Lett.}\ }%
  \textbf{\bibinfo {volume} {78}},\ \bibinfo {pages} {2940} (\bibinfo {year}
  {1997})%
  \bibAnnoteFile{NoStop}{Sirko1997}%
\bibitem{Kohler1997}%
  \BibitemOpen
  \bibfield{author}{%
  \bibinfo {author} {\bibfnamefont{A.}~\bibnamefont{Kohler}}, \bibinfo {author}
  {\bibfnamefont{G.~H.~M.}\ \bibnamefont{Killesreiter}},\ and\ \bibinfo
  {author} {\bibfnamefont{R.}~\bibnamefont{Bl\"{u}mel}},\ }%
  \bibfield{journal}{%
  \bibinfo {journal} {Phys. Rev. E}\ }%
  \textbf{\bibinfo {volume} {56}},\ \bibinfo {pages} {2691} (\bibinfo {year}
  {1997})%
  \bibAnnoteFile{NoStop}{Kohler1997}%
\bibitem{Bauch1998}%
  \BibitemOpen
  \bibfield{author}{%
  \bibinfo {author} {\bibfnamefont{S.}~\bibnamefont{Bauch}}, \bibinfo {author}
  {\bibfnamefont{A.}~\bibnamefont{Bledowski}}, \bibinfo {author}
  {\bibfnamefont{L.}~\bibnamefont{Sirko}}, \bibinfo {author}
  {\bibfnamefont{P.~M.}\ \bibnamefont{Koch}},\ and\ \bibinfo {author}
  {\bibfnamefont{R.}~\bibnamefont{Bl\"{u}mel}},\ }%
  \bibfield{journal}{%
  \bibinfo {journal} {Phys. Rev. E}\ }%
  \textbf{\bibinfo {volume} {57}},\ \bibinfo {pages} {304} (\bibinfo {year}
  {1998})%
  \bibAnnoteFile{NoStop}{Bauch1998}%
\bibitem{Schafer2001}%
  \BibitemOpen
  \bibfield{author}{%
  \bibinfo {author} {\bibfnamefont{R.}~\bibnamefont{Sch\"{a}fer}}, \bibinfo
  {author} {\bibfnamefont{U.}~\bibnamefont{Kuhl}}, \bibinfo {author}
  {\bibfnamefont{M.}~\bibnamefont{Barth}},\ and\ \bibinfo {author}
  {\bibfnamefont{H.-J.}\ \bibnamefont{St\"{o}ckmann}},\ }%
  \bibfield{journal}{%
  \bibinfo {journal} {Found. Phys.}\ }%
  \textbf{\bibinfo {volume} {31}},\ \bibinfo {pages} {475} (\bibinfo {year}
  {2001})%
  \bibAnnoteFile{NoStop}{Schafer2001}%
\bibitem{Oerter1996}%
  \BibitemOpen
  \bibfield{author}{%
  \bibinfo {author} {\bibfnamefont{R.~N.}\ \bibnamefont{Oerter}}, \bibinfo
  {author} {\bibfnamefont{E.}~\bibnamefont{Ott}}, \bibinfo {author}
  {\bibfnamefont{J.}~\bibnamefont{T.~M.~Antonsen}},\ and\ \bibinfo {author}
  {\bibfnamefont{P.}~\bibnamefont{So}},\ }%
  \bibfield{journal}{%
  \bibinfo {journal} {Phys. Lett. A}\ }%
  \textbf{\bibinfo {volume} {216}},\ \bibinfo {pages} {59} (\bibinfo {year}
  {1996})%
  \bibAnnoteFile{NoStop}{Oerter1996}%
\bibitem{Timberlake2009}%
  \BibitemOpen
  \bibfield{author}{%
  \bibinfo {author} {\bibfnamefont{T.~K.}\ \bibnamefont{Timberlake}}\ and\
  \bibinfo {author} {\bibfnamefont{M.~M.}\ \bibnamefont{Nelson}},\ }%
  \bibfield{journal}{%
  \bibinfo {journal} {Phys. Rev. E}\ }%
  \textbf{\bibinfo {volume} {79}},\ \bibinfo {pages} {036213} (\bibinfo {year}
  {2009})%
  \bibAnnoteFile{NoStop}{Timberlake2009}%
\bibitem{Doncheski2000}%
  \BibitemOpen
  \bibfield{author}{%
  \bibinfo {author} {\bibfnamefont{M.~A.}\ \bibnamefont{Doncheski}}\ and\
  \bibinfo {author} {\bibfnamefont{R.~W.}\ \bibnamefont{Robinett}},\ }%
  \bibfield{journal}{%
  \bibinfo {journal} {Eur. J. Phys.}\ }%
  \textbf{\bibinfo {volume} {21}},\ \bibinfo {pages} {217} (\bibinfo {year}
  {2000})%
  \bibAnnoteFile{NoStop}{Doncheski2000}%
\bibitem{Gilbert2005}%
  \BibitemOpen
  \bibfield{author}{%
  \bibinfo {author} {\bibfnamefont{L.~P.}\ \bibnamefont{Gilbert}}, \bibinfo
  {author} {\bibfnamefont{M.}~\bibnamefont{Belloni}}, \bibinfo {author}
  {\bibfnamefont{M.~A.}\ \bibnamefont{Doncheski}},\ and\ \bibinfo {author}
  {\bibfnamefont{R.~W.}\ \bibnamefont{Robinett}},\ }%
  \bibfield{journal}{%
  \bibinfo {journal} {Eur. J. Phys.}\ }%
  \textbf{\bibinfo {volume} {26}},\ \bibinfo {pages} {815} (\bibinfo {year}
  {2005})%
  \bibAnnoteFile{NoStop}{Gilbert2005}%
\bibitem{Jensen2005}%
  \BibitemOpen
  \bibfield{author}{%
  \bibinfo {author} {\bibfnamefont{Y.}~\bibnamefont{Dabaghian}}\ and\ \bibinfo
  {author} {\bibfnamefont{R.}~\bibnamefont{Jensen}},\ }%
  \bibfield{journal}{%
  \bibinfo {journal} {Eur. J. Phys.}\ }%
  \textbf{\bibinfo {volume} {26}},\ \bibinfo {pages} {423} (\bibinfo {year}
  {2005})%
  \bibAnnoteFile{NoStop}{Jensen2005}%
\bibitem{Griffiths2005}%
  \BibitemOpen
  \bibfield{author}{%
  \bibinfo {author} {\bibfnamefont{D.~J.}\ \bibnamefont{Griffiths}},\ }%
  \emph{\bibinfo {title} {Introduction to Quantum Mechanics}},\ \bibinfo
  {edition} {2nd}\ ed.\ (\bibinfo {publisher} {Pearson},\ \bibinfo {address}
  {Upper Saddle River, NJ},\ \bibinfo {year} {2005})%
  \bibAnnoteFile{NoStop}{Griffiths2005}%
\bibitem{Merzbacher1970}%
  \BibitemOpen
  \bibfield{author}{%
  \bibinfo {author} {\bibfnamefont{E.}~\bibnamefont{Merzbacher}},\ }%
  \emph{\bibinfo {title} {Quantum Mechanics}},\ \bibinfo {edition} {2nd}\ ed.\
  (\bibinfo {publisher} {John Wiley and Sons},\ \bibinfo {address} {New York},\
  \bibinfo {year} {1970})%
  \bibAnnoteFile{NoStop}{Merzbacher1970}%
\bibitem{BlumelComment}%
  \BibitemOpen
  \bibfield{author}{%
  \bibinfo {author} {\bibfnamefont{R.}~\bibnamefont{Bl\"{u}mel}},\ }%
  \bibfield{journal}{%
  \bibinfo {journal} {Eur. J. Phys.}\ }%
  \textbf{\bibinfo {volume} {27}},\ \bibinfo {pages} {L1} (\bibinfo {year}
  {2006})%
  \bibAnnoteFile{NoStop}{BlumelComment}%
\bibitem{numrec}%
  \BibitemOpen
  \bibfield{author}{%
  \bibinfo {author} {\bibfnamefont{W.~H.}\ \bibnamefont{Press}}, \bibinfo
  {author} {\bibfnamefont{S.~A.}\ \bibnamefont{Teukolsky}}, \bibinfo {author}
  {\bibfnamefont{W.~T.}\ \bibnamefont{Vetterling}},\ and\ \bibinfo {author}
  {\bibfnamefont{B.~P.}\ \bibnamefont{Flannery}},\ }%
  \emph{\bibinfo {title} {Numerical Recipes in FORTRAN 77}},\ \bibinfo
  {edition} {2nd}\ ed.\ (\bibinfo {publisher} {Cambridge UP},\ \bibinfo
  {address} {Cambridge},\ \bibinfo {year} {1992})%
  \bibAnnoteFile{NoStop}{numrec}%
\bibitem{DalgarnoLewis}%
  \BibitemOpen
  \bibfield{author}{%
  \bibinfo {author} {\bibfnamefont{H.~A.}\ \bibnamefont{Mavromatis}},\ }%
  \bibfield{journal}{%
  \bibinfo {journal} {Am. J. Phys.}\ }%
  \textbf{\bibinfo {volume} {59}},\ \bibinfo {pages} {738} (\bibinfo {year} {1991})%
  \bibAnnoteFile{NoStop}{DalgarnoLewis}%
\bibitem{LogPT}%
  \BibitemOpen
  \bibfield{author}{%
  \bibinfo {author} {\bibfnamefont{T.}~\bibnamefont{Imbo}}\ and\ \bibinfo
  {author} {\bibfnamefont{U.}~\bibnamefont{Sukhatme}},\ }%
  \bibfield{journal}{%
  \bibinfo {journal} {Am. J. of Phys.}\ }%
  \textbf{\bibinfo {volume} {52}},\ \bibinfo {pages} {140} (\bibinfo {year}
  {1984})%
  \bibAnnoteFile{NoStop}{LogPT}%
\end{thebibliography}
 \end{document}